\documentclass[aps,prl,twocolumn,floats,amsmath,amssymb,showpacs,floatfix]%
{revtex4}

\usepackage{graphicx}
\usepackage{dcolumn}
\usepackage{psfrag}
\renewcommand{\bbox}[1]{\mbox{\boldmath $#1$}}

\begin{document}
\title{Advective collisions}
\author{B. Andersson$^{1}$, K. Gustavsson$^{1}$,
B. Mehlig$^{1}$,
and M. Wilkinson$^{2}$}
\affiliation{
$^{1}$Department of Physics, G\"oteborg University, 41296
Gothenburg, Sweden \\$^{2}$Faculty of Mathematics and Computing, The Open
University, Walton Hall,
Milton Keynes, MK7 6AA, England \\}

\begin{abstract}
Small particles advected in a fluid can collide (and therefore
aggregate) due to the stretching or shearing of fluid elements.
This effect is usually discussed in terms of a theory due to
Saffman and Turner ({\it J. Fluid Mech.}, {\bf 1}, 16-30, (1956)).
We show that in complex or random flows the Saffman-Turner theory
for the collision rate describes only an initial transient (which
we evaluate exactly). We obtain precise expressions for the
steady-state collision rate for flows with small Kubo number,
including the influence of fractal clustering on the collision
rate for compressible flows. For incompressible turbulent flows,
where the Kubo number is of order unity, the Saffman-Turner
theory is an upper bound.
\end{abstract}

\maketitle

1. {\sl Introduction}. Suspensions of small particles in a fluid,
such as aerosols or colloids, are ubiquitous in the natural world
and in technology. Such systems may be unstable due to collisions
of the suspended particles giving rise to aggregation or to
chemical reaction. These collision processes are fundamental to
understanding the formation of raindrops from clouds \cite{Sha03},
or the growth of planets from dust in a protostellar environment
\cite{Arm07}.
\begin{figure}
\includegraphics[width=7cm,clip]{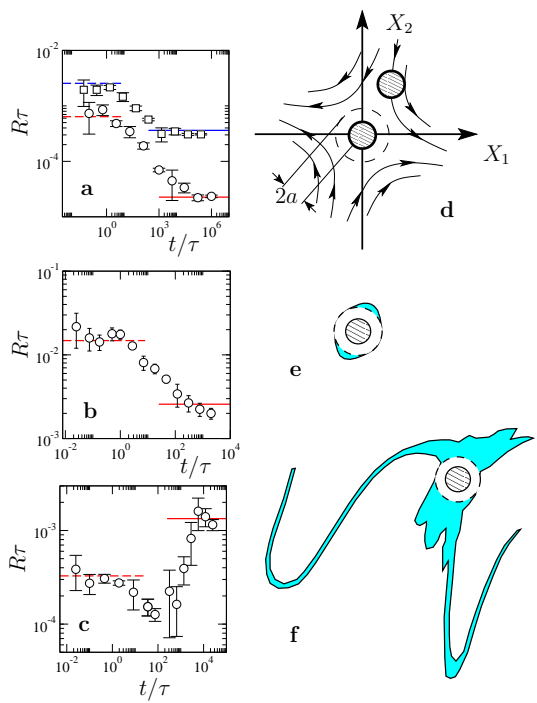}
\caption{\label{fig:1} Collision rate as a function of time for
particles advected in an random flow. {\bf a} incompressible flow
in two spatial dimensions, $u_0 = 1$, $\eta = 0.1$,
$n_0 = 10^3$, $\tau = 0.001$ ($\circ$) and $\tau = 0.004$
($\square$), $a=0.003$. Theory for initial transient (\ref{eq:
14}), dashed lines, for collision rate (\ref{eq:11}), full lines.
Error bars correspond to one standard deviation. {\bf b} three
spatial dimensions, $u_0 = 1$, $\tau = 0.004$, $\eta =
0.1$, $a=0.02$, and $n_0 = 2\times 10^3$. Theory for initial
transient (\ref{eq: 17}), for collision rate (\ref{eq:12}).  {\bf
c} two-dimensional compressible flow. Parameters: $u_0 = 1$, $\eta
= 0.1$, $\tau = 0.004$, $a=0.001$, $n_0 = 10^3$, and $\Gamma = 2$.
Initial transient (\ref{eq:it}), dashed line; colllision rate
(\ref{eq:23}), solid line.
 {\bf d} steady hyperbolic flow in the vicinity of a test particle,
 {\bf e} shows region (in rest frame of test particle)
 occupied by particles which will collide with test particle after time
 $t=20\tau$ (parameters as in Fig. 1{\bf a}, $\Box$), {\bf f} same
 but after time $t = 10^3\tau$.}
\end{figure}

Collisions between particles will always occur due to Brownian
diffusion. The rate of collision may be greatly increased if the
fluid is in macroscopic motion \cite{Smo17}, such as turbulence
\cite{Saf56}. If the inertia of the suspended particles is
significant, the suspended particles may move relative to the
fluid and this effect can greatly increase the rate of collisions
\cite{Abr75,Fal02,Wil06}. However if the particles are
sufficiently small that the effect of their inertia is negligible
(advective limit), the particles may be brought into contact by
the effect of shearing motion in the flow. This process was
discussed by Saffmann and Turner \cite{Saf56} and their theory has
been the starting point for most subsequent work on advective
collisions in smooth flows. Here we point out a previously
unremarked assumption in \cite{Saf56}, which implies that their
expression for the collision rate describes a short-lived
transient. In the case of incompressible flow their expression is
an upper bound: we give the first precise results on the
steady-state collision rate of their model. Both the
Saffman-Turner theory and our expression are compared with
numerical simulations in Fig. \ref{fig:1}{\bf a},{\bf b} (for
synthetic incompressible flows in two and three dimensions respectively; the
details are explained later). In both cases we observe an initial
transient given by the Saffman-Turner theory, and
at larger times the collision rate approaches a significantly
lower value which is in agreement with our theory.

Reference \cite{Saf56} treats the local flow in the vicinity of a
particle as if it were a steady hyperbolic flow, whereas in
reality it must fluctuate. We argue that the appropriate theory
for the collision rate must depend upon a dimensionless parameter
termed the Kubo number, ${\rm Ku}$.

2. {\sl Summary}. In this letter we give the first precise
expressions for advective collisions in smooth flows. We obtain
asymptotic results in the case of flows with small Kubo number, in
both two and three dimensions, including flows with Richardson
diffusion and compressible flows where particles cluster onto a
fractal set. We also describe new results on the evaluation of the
Saffman-Turner expression for the initial collision rate in two
and three dimensions.

3. {\sl A formula for the collision rate}. We start by discussing a
general formula for the collision rate, which is very similar to
that proposed by Saffmann and Turner \cite{Saf56}. We model
processes where particles coalesce upon contact with probability
unity. The particles collide when their separation falls below
$2a$ (where $a$ is the particle radius). For simplicity we
initially discuss two spatial dimensions. Let $v_r(r,\theta,t)$ be
the radial component of the velocity of the fluid relative to a
test particle at $(r,\theta)$ (in polar coordinates). We make the
approximation that the velocity field is not perturbed by the
suspended particles: this is valid in the Epstein low-pressure
regime \cite{Eps24} and is a reasonable approximation \cite{Saf56}
in conventional fluids. The collision rate for a single
particle is then the rate at which points distributed randomly
with density $n_0$ cross into the disc of radius $2a$ centered on
the origin. The instantaneous collision rate may then be written
as
\begin{equation}
\label{eq: 1} R\!=\!-\!2a n_0 \!\! \int_0^{2\pi}\!\!\!\!\!\!{\rm d}\theta
v_r(2a,\theta,t)\Theta(-v_r(2a,\theta,t))\chi(2a,\theta,t)
\end{equation}
where the Heaviside function $\Theta(\cdot)$ selects for particles
entering a disc of radius $2a$, $n_0$ is the number density of the
suspended particles in the vicinity of the test particle, and
$\chi(r,\theta,t)$ is zero if the fluid at $(r,\theta)$ at time
$t$ has previously passed through the disc of radius $2a$, and
unity otherwise. The collision rate of a single particle $R_{\rm
a}$ is the time-average of (\ref{eq: 1}), yielding an exact and
general expression for the advective collision rate, but in the
general case it is difficult to evaluate since   the
characteristic function $\chi$ depends on the history of the flow.

The relative speed $v_r$ is determined as follows \cite{Saf56}:
for small particles $a\ll \eta$ we approximate the flow
$u(\bbox{r},0)$ in the vicinity of the test particle by a
linearised equation for the small distance $\mbox{\boldmath$X$}$
from the test particle: $\dot{\mbox{\boldmath$X$}}= {\bf
A}(\bbox{0},0)\mbox{\boldmath$X$}$, where ${\bf A}(\bbox{r},t)$ is
the strain matrix of the flow $\bbox{u}(\bbox{r},t)$.
We then have $v_r=\dot{\mbox{\boldmath$X$}}\cdot \hat{\bf n}(\theta)$
where $\hat{\bf n}(\theta)$ is a unit vector normal to the surface
at a point with polar angle $\theta$. At the point
$\mbox{\boldmath$X$}=2a\hat{\bf n}(\theta)$ on the surface of the
disc
\begin{equation}
\label{eq: 2a}
v_r=2a\hat{\bf n}\cdot {\bf A}\hat{\bf n}\,.
\end{equation}
Eqs (\ref{eq: 1},\ref{eq: 2a}) appear as the key expressions for
the collision rate in \cite{Saf56}, but without the factor $\chi$.
Ref. \cite{Saf56} describes the flow in the vicinity of a test
particle as if it were a steady hyperbolic flow (Fig.
\ref{fig:1}{\bf d}). In that case, particles advected by the local
flow only pass through the circle $r=2a$ once (Fig.
\ref{fig:1}{\bf e}), so that the factor $\chi$ has no effect. In
general, however, the flow in the vicinity of the test particle
will fluctuate, as illustrated in Fig. \ref{fig:1}{\bf f}, so that
the directions of the eigenvectors of ${\bf A}$ rotate. The
function $\chi(r,\theta,t)$ equals zero for regions where the flow
at $r=2a$ has reversed direction, so that points that were
originally advected away return to the surface. This reduces the
collision rate below the Saffman-Turner estimate.

There is one limit in which the Saffman-Turner theory is exact:
this is the limit of very short times, when the initial condition
ensures that $\chi(r,\theta,0)=1$. For later times, it might be
expected that the Saffman-Turner theory is an upper bound on the
collision rate. This is true if the flow field is incompressible.
For compressible flows there is a complication, because particles
may cluster together, and the density in the vicinity of a test
particle may be larger than the average density, $n_0$.

Although it is not possible to make a precise theory for the
steady-state collision rate in the general case, this can be done
in the limit where the velocity field of the fluid is rapidly
fluctuating. The effect of fluctuations depends on a dimensionless
parameter termed the Kubo number. A random or turbulent flow may
be characterised by a correlation length $\eta$, a correlation
time $\tau$, and a characteristic velocity $u_0$. The Kubo number
${\rm Ku}=u_0\tau/\eta$ is a dimensionless measure of how rapidly
the flow fluctuates. It might be argued that the Saffman-Turner
theory, which treats the flow as if it were steady, would apply to
flows with very large Kubo number. However, such flows are
unphysical, because the velocity field transports spatial
fluctuations so that they become temporal fluctuations, implying
that the Kubo number is never large. In the case of
fully-developed turbulence, ${\rm Ku}=O(1)$.

4. {\sl Collision rate for small Kubo number}. In the limit Ku
$\to 0$ (that is, where the correlation time approaches zero), the
relative separation of two particles undergoes a diffusion process
determining the collision rate.

We consider the case of two spatial dimensions.
By homogeneity we can
assume that the particles are at ${\bf 0}$ and
$\mbox{\boldmath$X$}=(X_1,X_2)$
and by isotropy we can take $X_2=0$ and write $\mbox{\boldmath$X$}=(r,0)$.
The
fluctuation $\delta X_i$ of the separation in a short time $\delta
t$ satisfies $\langle \delta X_i\rangle=0$. The correlation of
these increments is $\langle \delta X_i\delta
X_j\rangle=2D_{ij}\delta t$, with
\begin{equation}
\label{eq: 2}
D_{ij}\!=\!{1\over 2}\int_{-\infty}^\infty
\!\!\!\!\!\!{\rm d}t \,\langle
[u_i(\mbox{\boldmath$X$,t})\!-\!u_i({\bf
0},t)][u_j(\mbox{\boldmath$X$},0)\!-\!u_j({\bf 0},0)]\rangle\,.
\end{equation}
These \lq diffusion constants' are computed as follows. We
consider a velocity field $\mbox{\boldmath$u$}(\bbox{r},t)$
generated by a \lq stream function' $\psi(\bbox{r},t)$:
$\bbox{u} = \bbox{\nabla}\wedge \psi \hat{\bf n}_z$, where
$\hat{\bf n}_z$ is a unit vector in the $z$-direction,
with correlation function
$\langle \psi(\mbox{\boldmath$r$},t)\psi({\bf
0},0)\rangle=C(r,t)$, where $r=\vert\mbox{\boldmath$r$}\vert$.
For $\mbox{\boldmath$X$}=(r,0)$ we find
$D_{12}=D_{21}=0$ and
\begin{eqnarray}
\label{eq: 3} D_{11}(r)&=&-\int_{-\infty}^\infty{\rm d}t\,[
C''(0,t)-{1\over
r}C'(r,t)]\nonumber \\
D_{22}(r)&=&-\int_{-\infty}^\infty {\rm d}t\,[ C''(0,t)-C''(r,t)]
\end{eqnarray}
where the prime denotes a derivative w.r.t. $r$.

We set up a Fokker-Planck equation for the separation $r$ of two
particles. The change in the value of $r$ in $\delta t$ is $\delta
r=\sqrt {(x+\delta x)^2+\delta y^2}-x=\delta x+\delta y^2/(2x)\,,$
so that $\langle \delta r\rangle=D_{22}(r)\delta t/r$ and $\langle
\delta r^2\rangle = \langle \delta x^2\rangle=2D_{11}(r) \delta
t$. We obtain the Fokker-Planck equation
\begin{equation}
\label{eq:fp} {\partial\rho\over{\partial t}}=
{\partial\over{\partial r}} \biggl[-{D_{22}(r)\rho\over{
r}}+{\partial\over{\partial r}} \bigl(D_{11}(r)\rho\bigr)\biggr]\
.
\end{equation}
The collision rate is determined by a steady-state solution of
(\ref{eq:fp}) with a constant flux $-J$. The boundary condition is
$\rho(2a)=0$, which corresponds to particles which reach $r=2a$
being absorbed, so that there are no re-collisions. The collision
rate is $R_{\rm a}=J$. To enforce the boundary condition at
$r=2a$, it is sufficient to consider an approximate solution valid
at small values of $r$. Expanding the diffusion constants for
small $r$ we find $D_{11}(r)\sim {\cal D} r^2$ and $D_{22}(r) \sim
3 {\cal D} r^2$, with
\begin{equation}
\label{eq:calD} {\cal D} = \frac{1}{6}\int_{-\infty}^\infty
\!\!\!\!{\rm d}t\, C''''(0,t)
\end{equation}
we obtain  the desired steady-state solution
\begin{equation}
\label{eq:9}
\rho(r)=J/(2 {\cal D} r)+A r\,.
\end{equation}
To determine the constant $A$ we match
the solution (\ref{eq:9})
for $J=0$ to the exact steady-state zero-flux solution of (\ref{eq:fp}).
The latter satisfies
\begin{equation}
\label{eq: 10} {\partial \log \rho\over {\partial r}}-{1\over
r}={1\over {D_{11}}}\biggl[{D_{22}\over r}-{\partial
D_{11}\over{\partial r}}\biggr]-{1\over r}\equiv \alpha(r)\ .
\end{equation}
From (\ref{eq: 3}) we find $\alpha(r)=0$. The density is thus
proportional to $r$, normalisation gives $\rho(r) =
2\pi n_0 r$. We conclude that $A=2\pi n_0$ in (\ref{eq:9}). The
collision rate is determined by setting $\rho(2a)=0$ in
(\ref{eq:9}) and solving for $R_{\rm a}=J$:
\begin{equation}
\label{eq:11}
R_a = 16 \pi \,{\cal D}\,n_0  a^2\,.
\end{equation}

5. {\sl Richardson diffusion}. In a steady, fully developed
turbulent flow \cite{Fri97}, particle separations exhibit
Richardson diffusion ${\rm d} \langle r^2\rangle /{\rm d}t \propto
r^{4/3}$ in the inertial range $r \gg \eta$ \cite{Ric26} (where
$\eta$ is the Kolmogorov length). This implies $D_{ij}(r) \propto
r^{4/3}$. It is of interest to consider the collision rate in a flow
exhibiting Richardson diffusion in the limit ${\rm Ku}\to 0$,
to establish whether additional complications arise when
considering a multiscale flow (although real turbulence has ${\rm
Ku}=O(1)$, and is at the borderline of applicability of our
theory). In the following we argue that (\ref{eq:11}) is unchanged
for a flow exhibiting Richardson diffusion, such that
$D_{ij}(r)\sim r^\xi$ in the inertial range.

At first sight, given eq. (\ref{eq:fp}), this power-law scaling of
the diffusion constants suggests that there should be a power-law
behaviour of $\rho(r)$ in the inertial range, with an exponent
which depends upon $\xi$. This would be inconsistent with the
requirement that $\rho(r)\sim 2\pi n_0 r$ for $r\gg \eta$ (as it
must be in a homogenous and isotropic flow). However, given
equations (\ref{eq: 3}), we see that $C(r,t)$ contains a term
which can be approximated by a multiple of $r^{2+\xi}$ for $r$ in
the inertial range. It follows that the coefficients of $D_{11}$
and $D_{22}$ are related, $D_{11}(r) = C r^\xi$ and $D_{22}(r) =
C(1+\xi) r^\xi$. Comparison with (\ref{eq:fp}) shows that
$\rho\sim r$ in the inertial range, as expected. Since the
velocity field in two dimensions is generated by a stream
function, it follows immediately that $\alpha(r)= 0$, and thus
$\rho(r) = 2\pi n_0 r$ for $r\ll \eta$ as well as for $r\gg \eta$.
We conclude that our result (\ref{eq:11}) for the collision rate
of particles advected in incompressible flows at small values of
${\rm Ku}$ is insensitive to what happens in the inertial range.

6. {\sl Computer simulations}. The numerical simulations shown in
Figs. \ref{fig:1}{\bf a} and {\bf b} employ a Gaussian random
stream function with spatially and temporally stationary,
spatially isotropic statistics with mean value zero and $C(r,t) =
({u_0^2 \eta^2}/2)\, \exp[-|t|/\tau-r^2/(2\eta^2)]$, for which
${\cal D} = u_0^2\tau/(2\eta^2)$. The particles are initially
randomly positioned, and they are regarded as having collided when
their separation falls below $2a$. Our plots show the rate of
collision with a test particle which is also advected by the
velocity field. Upon collision with the test particle, colliding
particles are removed. The collision rate was determined by
dividing by $t$ the  number of particles which collided in the
time interval $[0,t]$.
This yields accurate results for the initial transient, $R_0$,
and for the steady-state collision rate, $R_{\rm a}$.
Only a
small fraction of particles had collided by the end of each run
(less than  a few percent). Fig. \ref{fig:1}{\bf a} shows the
collision rate  as a function of time for particles advected in a
two-dimensional incompressible flow for small Kubo number. We
observe good agreement with the analytical result (\ref{eq:11}),
full line. In the following we show how to derive the initial
transients shown in Fig. \ref{fig:1}{\bf a}, dashed lines.

7. {\sl Exact result for the initial transient}. The initial
transient is obtained by averaging equations (\ref{eq: 1}),
(\ref{eq: 2a}) over the ensemble of random strain matrices
determined by the statistics of the stream function
$\psi(\bbox{0},0)$. In order to compute the average of $R$, we
decompose the matrix ${\bf A}$ which appears in (\ref{eq: 2a})
into a symmetric ${\bf A}_{()}$ and an antisymmetric ${\bf
A}_{[]}$ part and
 sort the eigenvalues $\sigma_i, i=1,\ldots,d$ of the symmetric part
in order: $\sigma_1\le \sigma_2\le,\ldots,\le \sigma_d$.
In two dimensions we obtain for a given ${\bf A}$
\begin{align}
R(\sigma_2) &= 8 n_0a^2\sigma_2\,.
\end{align}
We average over the distribution of the largest eigenvalue of
${\bf A}_{()}$, $P(\sigma_2)=(2\sigma_2 \eta^2/u_0^2)
\exp[-\sigma_2^2\eta^2/u_0^2]$, where $\sigma_2\ge 0$, and find
for the initial rate of collision, $R_0$:
\begin{align}
\label{eq: 14} R_0= 4\sqrt{\pi} \,n_0 a^2\,u_0/\eta\,.
\end{align}
This initial transient is shown in Fig. \ref{fig:1} {\bf a} (dashed
lines). Comparing with (\ref{eq:11}), we have $R_{\rm
a}\sim {\rm Ku}R_0$, demonstrating that the effect of eliminating multiple
collisions is significant when the velocity field $\bbox{u}(\bbox{r},t)$
is rapidly
fluctuating.

8. {\sl Collision rate in the presence of fractal clustering}.
When the velocity field has a compressible component, the
particles cluster onto a fractal set \cite{Som93}. This may alter
the collision rate, because particles which cluster together are
expected to collide more frequently \cite{Fal02}. Here we present
the first exact results on this effect. When considering
compressibility effects we restrict ourselves to two spatial
dimensions, because the natural experimental realisation of
particles advected by compressible flows involves particles
floating on the surface of a turbulent fluid, such as the
experiments discussed in \cite{Som93,Cre04}.

We consider a partially compressible flow of the form $\bbox{u} =
(\bbox{\nabla}\wedge \psi \hat {\bf n}_z +
\beta\bbox{\nabla}\phi)/\sqrt{1+\beta^2}$ where $\psi$ and $\phi$
are independent Gaussian random functions with correlation
function $C(r,t)$ and $\beta>0$. As before we seek a
solution of (\ref{eq:fp}) with constant flux $-J$. We have
\begin{eqnarray}
\label{eq:21}
\mbox{}\hspace*{-4mm}D_{11}(r)\!\!\!%
&=&\!\!-\!\!\int_{-\infty}^\infty\!\!\!\!\!\!\!\!{\rm d}t\, [
C''(0,t)\!-\!{C'(r,t)\over r(1\!+\!\beta^2)}\!-\!
\frac{\beta^2 C''(r,t)}{1\!+\!\beta^2}]\,\\
D_{22}(r)\!\!\!&=&\!\!-\!\!\int_{-\infty}^\infty
\!\!\!\!\!\!\!\!{\rm d}t \,
[C''(0,t)\!-\!\frac{C''(r,t)}{1\!+\!\beta^2}\!-\!{\beta^2C'(r,t)\over
r(1\!+\!\beta^2)}]\,\nonumber
\end{eqnarray}
so that $D_{22}/D_{11} \rightarrow
\Gamma \equiv(\beta^2+3)/(3\beta^2+1)$ as $r \to 0$.
The steady-state solution of (\ref{eq:fp})
is for $r \ll \eta$
\begin{equation}
\label{eq: 21} \rho(r) =[(\Gamma\!+\!1)/(\Gamma\!-\!1)]\, J/(4
{\cal D} r) + A r^{\Gamma-2}\,.
\end{equation}
The constant $A$ is determined, as before, by matching to the
small-$r$ behaviour of the exact zero-flux solution of (\ref{eq:
10}) with the diffusion constants (\ref{eq:21}):
\begin {equation}
\label{eq: 23}
\rho(r) = 2\pi n_0 r
\exp\left[ -\int_r^\infty\!\!\!\! {\rm d}r' \alpha(r')
\right]\,.
\end{equation}
In order to match (\ref{eq: 21}) with (\ref{eq: 23}) we evaluate
(\ref{eq: 23}) for small values of $r$. One obtains
$\rho(r) \sim 2 \pi n_0 \eta (r/\eta)^{D_2-1}$ where $D_2 = \Gamma-1$
is the correlation dimension of the fractal set onto which the particles
cluster (this dimension was first determined in Ref. \cite{Fal99}).
Matching with (\ref{eq: 21}) we obtain (for $1 < \Gamma \leq 3$)
\begin{equation}
\label{eq:23} R_{\rm a} = [({\Gamma-1})/({\Gamma+1})] \, 8\pi
{\cal D}\,\,n_0\eta^2\,(2a/\eta)^{D_2}\,.
\end{equation}
For the initial transient in compressible flows we find:
\begin{equation}
\label{eq:it} R_0 =
 (8\sqrt{\pi}/\sqrt{1+\Gamma})\,
 n_0 a^2\, {u_0}/{\eta}\,.
\end{equation}
Comparison with (\ref{eq:23}) shows that eq. (\ref{eq: 1}) no
longer gives an upper bound of the collision rate upon setting
$\chi=1$. This is because Eqs. (\ref{eq: 1}) and (\ref{eq: 14})
assume a uniform distribution of particles. But for times $t \gg
|\lambda_1+\lambda_2|^{-1}$ (here $\lambda_1 > \lambda_2$ are the
Lyapunov exponents of the flow), the particles cluster on a
fractal set \cite{Som93} which can lead to an increase in the
collision rate (Fig. 1{\bf c}). 

There is a critical point at
$\Gamma = 1$ where $\lambda_1$ becomes negative and
path-coalescence occurs \cite{Meh04}. Experimental evidence
indicates that $\Gamma=1$ for surface flows above a turbulent
fluid \cite{Cre04}. At this critical point we obtain $R_{\rm
a}=4\pi {\cal D}\,n_0\eta^2\,\log(2a/\eta)$.

9. {\sl Three spatial dimensions}. We consider a three-dimensional
incompressible flow generated by three random vector
fields \cite{Dun05}. The calculation of $R_{\rm
a}$ proceeds in a fashion similar to the derivation of
(\ref{eq:11}):
 \begin{equation}
 \label{eq:12}
 R_{\rm a} =96\pi \,{\cal D}\,n_0 a^3 \ ,\ \ \ \ \
 {\cal D} = \frac{1}{2}{{\rm d}^2D_{11}(r)\over{{\rm d}r^2}}\bigg\vert_{r=0}\ .
 \end{equation}
As in the two-dimensional case we find that Richardson diffusion
does not alter the result (\ref{eq:12}). The initial transient in
three dimensions is derived as outlined above. We have obtained an
explicit expression for $R(\sigma_2, \sigma_3$) in terms of the
largest eigenvalues $\sigma_2,\sigma_3$ of ${\bf A}_{()}$, but
could not perform the average of $\sigma_2,\sigma_3$ analytically.
Numerical averaging for our Gaussian-correlated model gives
\begin{equation}
\label{eq: 17} R_0 \approx 24\, n_0 a^3\,u_0/\eta\,.
\end{equation}
Results (\ref{eq:12}) and (\ref{eq: 17})
are shown in Fig. \ref{fig:1}{\bf b} in comparison
with numerical simulations.

{\sl Acknowledgements.} We acknowledge support from
Vetenskapsr\aa{}det and from  \lq Nanoparticles in an interactive
environment' at G\"oteborg university.

\end{document}